\documentclass[twocolumn,pre,superscriptaddress]{revtex4}
\usepackage{latexsym}
\usepackage{graphicx}
\usepackage{epsfig,subfigure}
\begin{document}
\title{A food-web based unified model of ``macro''- and ``micro-'' evolution}

\author{Debashish Chowdhury{\footnote{E-mail: debch@iitk.ac.in}}} 

\affiliation{Department of Physics, Indian Institute of Technology, Kanpur 208016, India.}

\author{Dietrich Stauffer{\footnote{E-mail: stauffer@thp.uni-koeln.de}}}
\affiliation{Institute for Theoretical Physics, Cologne University, D-50923 K\"oln, Euroland.}

\begin{abstract}
We incorporate the generic hierarchical architecture of foodwebs into 
a ``{\it unified}'' model that describes both ``micro'' and ``macro'' 
evolutions within a single theoretical framework. This model describes 
the ``micro''-evolution in detail by accounting for the birth, ageing 
and natural death of individual organisms as well as prey-predator 
interactions on a hierarchical dynamic food web. It also provides a 
natural description of random mutations and speciation/orgination of 
species as well as their extinctions. The distribution of lifetimes of 
species follows an approximate power law only over a limited regime.

\end{abstract}

\maketitle
%Evolution | Monte Carlo simulation | Extinction | Lifetime distribution
\bigskip


\section{Introduction}

The questions of ``origin'' and ``evolution'' have always fascinated 
scientists in all disciplines. Physicists have focussed attention 
mostly on cosmological evolution and origin of universe. On the other 
hand, chemists and biologists have studied chemical evolution (i.e., 
formation of elements and compounds) as well as pre-biotic evolution 
and ``origin'' of life. Similarly, paleontologists try to understand 
the origin of species and evolution of eco-systems by reading ``history 
of life written on stone'' in the form of fossil records. In a recent 
Letter \cite{csk} we developed a dynamic network model for studying 
some generic features of the biological evolution of eco-systems. In 
this paper we extend that model incorporating the generic trophic-level 
architecture of food webs and show how it can account for evolution at 
both ecological as well as geological time scales.

\section{Earlier models and their limitations} 

Because of the close similarity between the evolution of interacting 
species and that of conventional systems of interacting agents 
studied in statistical physics, several models of ``macro''-evolution 
of eco-systems have been reported over the last decade in the physics 
literature (see \cite{drossel,newman,solerev} for recent reviews). 
Some of these describe macro-evolution as random walks on fitness 
landscape \cite{kauffman,bak} (see also \cite{peliti,wilke} for reviews), 
while some others have been formulated in terms of a matrix of 
inter-species interactions \cite{solerev,sole}. 
However, most of these models of ``macro''-evolution do not account for 
the dynamics of populations of species even in a collective manner. 
In other words, such models ignore biological details that are 
certainly important on {\it ecological} time scales and, therefore, 
cannot provide a natural description of origin, evolution and 
extinctions in terms of population dynamics.

On the other hand, the Lotka-Volterra equation \cite{lotka} has been 
used extensively in the mathematical modelling of population dynamics 
of prey-predator systems. However, for the study of population 
dynamics of entire ecological communities one needs a model of the 
food web \cite{cohenprs}. A food web \cite{foodweb,foodweb1,foodweb2} 
corresponding to an eco-system is a graphic description of prey-predator 
relations. More precisely, a food web is a directed graph where each 
node is labelled by a species' name and each directed link indicates 
the direction of flow of nutrient (i.e., {\it from} a prey {\it to} one 
of its predators).  However, most often, these models assume static 
food webs, where inter-species interactions are assumed to be independent 
of time. But, in real eco-systems, species are known to change their 
food habits with time \cite{thompson}. These changes in diets may be 
caused by scarcity of the normal food and abundance of alternative food 
resources. This may also arise from the adaptations of the prey species 
that tend to avoid being eaten by predators through camouflage or other 
mechanisms. Therefore, Lotka-Volterra type models with time-independent 
food webs cannot be expected to account for ``macro''-evolution of the 
eco-system over {\it geological} time scales.

Limitations of both these approches are well known \cite{mckane}, 
and attempts have been made to merge population dynamics and 
``macro''-evolution within a single mathematical framework \cite{per}.
Population dynamics is monitored in Abramson's ``macro''-evolutionary 
model \cite{abramson} in a simplified manner. However, Abramson 
postulated an oversimplified model of dynamically evolving food web 
that, essentially, consists of a single food chain.
Amaral and Meyer \cite{amaral} developed a ``macro''-evolutionary model 
with a dynamically evolving food web where niches are arranged in a 
hierarchical trophic level architecture. However, population dynamics 
of the species does not enter explicitly in this model. The strength 
of this model is its simplicity as some of its properties, e.g., its 
self-organized criticality, can be studied analytically 
\cite{drossel98,camacho00}. However, we feel, more details need to be 
included to address a wider range of biologically relevant questions.

\section{The ``unified'' eco-system model}

%%%%%%%%%%%%%%%%%%%%%%%%%%%%%%%%%%%%%%%%%%%%%%%%%%%%%%%%%%%%%%%%%%%%
\begin{figure}[tb]
\begin{center}
\includegraphics[angle=-90,width=0.9\columnwidth]{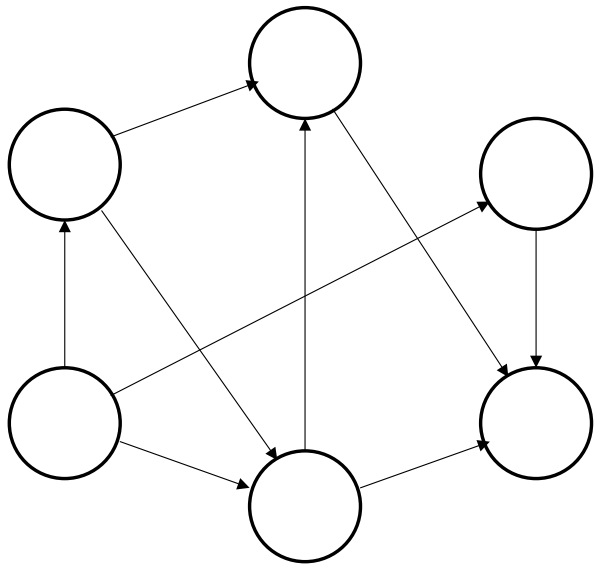}
\end{center}
\caption{A schematic representation of the network model, with {\it 
random} foodweb architecture, considered in \cite{csk}. The circles 
represent the niches in the eco-system. The arrows indicate the 
directions of nutrient flows {\it to} the species at an arbitrary 
stage during the evolution of the eco-system.}
\label{fig-0}
\end{figure}
%%%%%%%%%%%%%%%%%%%%%%%%%%%%%%%%%%%%%%%%%%%%%%%%%%%%%%%%%%%%%%%%%%%%

To our knowledge, our recent ``unified'' model \cite{csk} is one of the  
first few \cite{dudek,lasz} that describes not only ``macro''-evolution 
of origin/speciation and extinction of species on geological time scales 
but also ``micro''- evolutionary processes like, for example, the birth, 
growth (ageing) and natural death of individual organisms as well as the 
effects of prey-predator interactions on their populations. Our "unified" 
model, reported in \cite{csk}, can be schematically represented by the 
random network shown in figure \ref{fig-0}. Each node of this 
network, denoted by the circles, represents a niche that can be occupied 
by at most one species at a time. In that Letter \cite{csk} we postulated 
a simple random, but dynamic, food web ignoring the hierarchical 
organization of species in food webs. In this paper we postulate a generic 
hierarchical food web, where niches are arranged in different trophic 
levels, with biologically realistic inter-species interactions. 

\subsection{Architecture of the network} 

\begin{figure}[tb]
\begin{center}
\includegraphics[angle=-90,width=0.9\columnwidth]{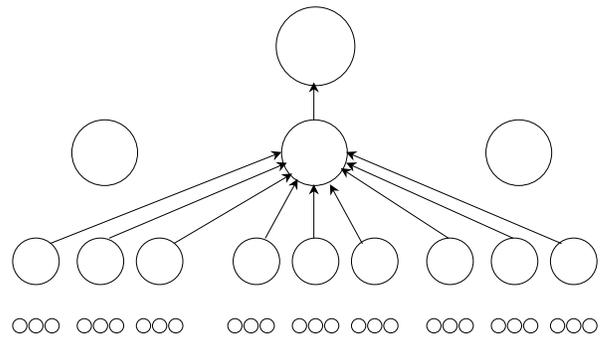}
\end{center}
\caption{A schematic representation of the network model, with {\it 
hierarchical} foodweb architecture. The circles represent the niches 
in the eco-system. Each arrow represents direction of nutrient flow. 
All possible nutrient flows  {\it to} the species occupying the second 
node at the second level and that occupying the highest level are 
shown explicitly.  }
\label{fig-1}
\end{figure}

As in our earlier work \cite{csk}, we model the eco-system as a dynamic 
{\it network} each node of which represents a niche that can be occupied 
by at most one species at a time. We assume a generic {\it hierarchical 
architecture} of this network (see fig.\ref{fig-1}) in order to capture 
the organization of species in different trophic levels of foodwebs 
\cite{foodweb}. If the $i$-th species occupies the $\nu$-th node at the 
${\ell}$-th trophic level of the food web, we denote its position by the 
ordered pair ${\ell}, \nu$. We assume only one single species at the 
highest level $\ell = 1$. Each node at level ${\ell}$ leads to $m$ 
branches at the level ${\ell} + 1$; therefore, the maximum allowed number 
of nodes in level ${\ell}$ is $m^{{\ell}-1}$ and the allowed range of 
${\ell}$ is $1 \leq {\ell} \leq {\ell}_{max}$. The hierarchical 
architecture helps us in capturing a well known fact that in the normal 
ecosystems the higher is the trophic level the fewer are the number of 
species.

%%%%%%%%%%%%%%%%%%%%%%%%%%%%%%%%%%%%%%%%%%%%%%%%%%%%%%%%%%%%%%%%%%%%
%%%%%%%%%%%%%%%%%%%%%%%%%%%%%%%%%%%%%%%%%%%%%%%%%%%%%%%%%%%%%%%%%%%%

\subsection{The network is dynamic}

The faster dynamics within each node captures "micro"-evolution, i.e., 
the birth, growth (ageing) and natural death of the individual organisms. 
Moreover, the network itself evolves slowly over sufficiently long time 
scales. For example, the adaptive evolution of the species takes place 
through alterations in some of their crucial characteristics by random 
mutations. 
Furthermore, as the eco-system evolves with time, the populations of 
some species would drop to zero, indicating their extinction, and the 
corresponding nodes would be slowly re-occupied by new species through the 
process of speciation.

At any arbitrary instant of time $t$ the model consists of $N(t)$ 
{\it species} each of which occupies one of the nodes of the dynamic 
network. The total number of species cannot exceed 
$N_{max} = (m^{{\ell}_{max}}-1)/(m-1)$, the total number of nodes.  
Our model allows $N(t)$ to fluctuate with time over the range 
${\ell} \leq N(t) \leq N_{max}$. The population (i.e., the total number 
of organisms) of a given species, say, $i$, at any arbitrary instant of 
time $t$ is given by $n_i(t)$. The {\it intra}-species interactions 
among the organisms of the same species for limited availability of 
resources, other than food, imposes an upper limit $n_{max}$ of the 
allowed population of each species. Thus, the total number of organisms 
$n(t)$ at time $t$ is given by $n(t) = \sum_{i=1}^{N(t)} n_i(t)$. Both 
$N_{max}$ and $n_{max}$ are time-independent parameters in the model.

\subsection{Interactions in the food web} 

Between any two species $i,k$ that occupy two adjacent trophic levels 
there is either a link ($J_{ik}=\pm1$) or no link ($J_{ik}=0$). The 
sign of $J_{ik}$ gives the direction of trophic flow, i.e. it is $+1$ 
if $i$ eats $k$ and it is $-1$ if $k$ eats $i$. Thus, $J_{ik} = 0$ 
means that there is no prey-predator relation between the two species 
$i$ and $k$. 

If we neglect parasites and herbivorous insects on trees, then, in 
general, predators are rarer and bigger than their prey \cite{cohen03}. 
This is very naturally incorporated in the hierarchical food web 
structure of our model by assuming that each predator needs $m$ prey 
animals to survive (see factor $m$ below).  The maximum number of 
individuals on each level $\ell$ is $m$ times bigger than on its 
predator level $\ell-1$ in the model, and when we imagine the predator
mass to be $m$ times the prey mass, then the maximum (and initial) amount 
of biomass on each level is the same.  In this way, the body size and 
abundance of a species are strongly correlated to the food web and its 
interactions with other species \cite{cohen93,cohen03}. 

The $J$ account not only for the {\it inter}-species interactions but 
also {\it intra}-species interactions. Let $S_i^+$ be the number of all 
prey individuals for species $i$ on the lower trophic level, and $S_i^-$ 
be $m$ times the number of all predator individuals on the higher 
trophic level. Since a predator eats $m$ prey per time interval, $S_i^+$ 
gives the available food for species $i$, and $S_i^-$ is the contribution 
of species $i$ to all predators on the higher level. If the available 
food $S_i^+$  is less than the requirement, then some organisms of the 
species $i$ will die of {\it starvation}, even if none of them is killed 
by any predator. This way the model can account not only 
for the inter-species  prey-predator interactions but also for the 
intra-species interactions arising from the competition of individual 
organisms during shortage of food supply.

Note that the food resources of a given species are not restricted to 
only the lower branches emanating from that node but it can also 
exploit the species at the lower-level nodes emanating from other 
nodes at its own trophic level. Moreover, note that although there is no 
direct interaction between species at the same trophic level in our model, 
they can compete, albeit indirectly, with each other for the same food 
resources available in the form of prey at the next lower trophic level.

\subsection{The collective characteristics of species}

An arbitrary species $i$, occupying the $\nu$-th node at the ${\ell}$-th 
level is {\it collectively} characterized by \cite{csk}:\\
(i) the {\it minimum reproduction age} $X_{rep}(i)$,\\ 
(ii) the {\it birth rate} $M(i)$,\\
(iii) the {\it maximum possible age} $X_{max}(i)$. \\ 
An individual of the species $i$ can reproduce only 
after attaining the age $X_{rep}(i)$. Whenever an organism of 
this species gives birth to offsprings, $M(i)$ of 
these are born simultaneously. None of the individuals of this
species can live longer than $X_{max}(i)$, 
even if an individual manages to escape its predators.

Note that, in several earlier works the reproductive success was modelled 
mathematically by assigning a ``fitness'' to a species or to an individual 
organism. The use of the term ``fitness'' has an interesting history 
\cite{brookfield}. In contrast to these earlier works, in our models, we 
assign a minimum reproductive age, a maximum possible age and the birth 
rate to model the reproductive success (or failure). It has been felt 
\cite{brookfield} that fitness merely summarizes, instead of explainig, 
the ability to survive and reproduce. On the other hand, the interplay 
of the $M, X_{rep}$ and $X_{max}$, we hope, will be able to explain why 
some species survive while others become extinct.

\subsection{The dynamics of the eco-system}

The state of the system is updated in discrete time steps as follows: 

\noindent {\it Step I- Birth}: Assuming, for the sake of simplicity, the 
reproduction to be {\it asexual}, each individual organism $\alpha$ 
($\alpha = 1,...,n_i(t)$) of the species $i$ ($i=1,2,...N(t)$) is 
allowed to give birth to $M(i;t)$ offsprings at every time step $t$ 
with probability (per unit time) $p_b(i,\alpha;t)$ which is non-zero 
only when the individual organism's age $X(i,\alpha;t) \ge X_{rep}(i;t)$. 

\noindent {\it Step II- Natural death}: At any arbitrary time step $t$ the 
probability (per unit time) of ``natural'' death (due to ageing) of 
an individual organism $\alpha$ of species $i$ is $p_d(i,\alpha;t)$. 

\noindent {\it Step III- Mutation}: With probability $p_{mut}$ per unit time, 
each of the species simultaneously increases or decreases, with equal 
probability, their $X_{rep}$, $X_{max}$ and $M$ by unity. (The ages are
restricted to the interval from 1 and 100, and $M > 0$.) Moreover, 
with the same probability $p_{mut}$ per unit time, they also re-adjust one of 
the links $J$ from prey and one of the links $J$ to predators \cite{sole}; 
if the link $J$ was zero, it is assigned a new value of $\pm 1$ 
whereas if the link was non-zero it is assigned a new 
value of zero. These re-adjustments of the incoming and outgoing 
(in the sense of nutrient flow)
interactions are intended to capture the facts that each species 
tries to minimize predators but look for new food resources.

\noindent {\it Step IV- Starvation death and killing by prey}: 
If $n_i-S_i^+$ is larger than $S_i^-$ 
then food 
shortage will be the dominant cause of premature death of a 
fraction of the existing population of the species $i$. On the 
other hand, if $S_i^- > n_i-S_i^+$, then a fraction 
of the existing population will be wiped out primarily by the 
predators. In order to capture these phenomena, at every time 
step $t$, in addition to the natural death due to ageing, a 
further reduction of the population by  
\begin{equation}
C ~~\max(S_i^-,~n_i- S_i^{+})
\label{eq-kill}
\end{equation}
is implemented where $n_i(t)$ is the population of the species $i$ 
that survives after the natural death step above. $C$ is a constant 
of proportionality. If implementation of these steps makes $n_i \leq 0$, 
species $i$ becomes extinct.  

\noindent {\it Step V- Speciation}: After the extinction of, typically, 
half of the species in a trophic level, the niches (nodes) left empty  
are re-filled by new species, with probability $p_{sp}$. All the 
simultaneously re-filled nodes in a trophic level of the network 
originate from {\it one common ancestor} which is picked up randomly 
from among the surviving species at the same trophic level. All the 
interactions $J$ of the new species are identical to those of their 
common ancestor. The characteristic parameters $X_{max}$, $X_{rep}$, 
$M$ of each of the new species differ randomly by $\pm 1$ from the 
corresponding parameters for their ancestor.  

%%%%%%%%%%%%%%%%%%%%%%%%%%%%%%%%%%%%%%%%%%%%%%%%%%%%%%%%%%%%%%%%%%%%
\begin{figure}[tb]
\begin{center}
\includegraphics[angle=-90,width=0.9\columnwidth]{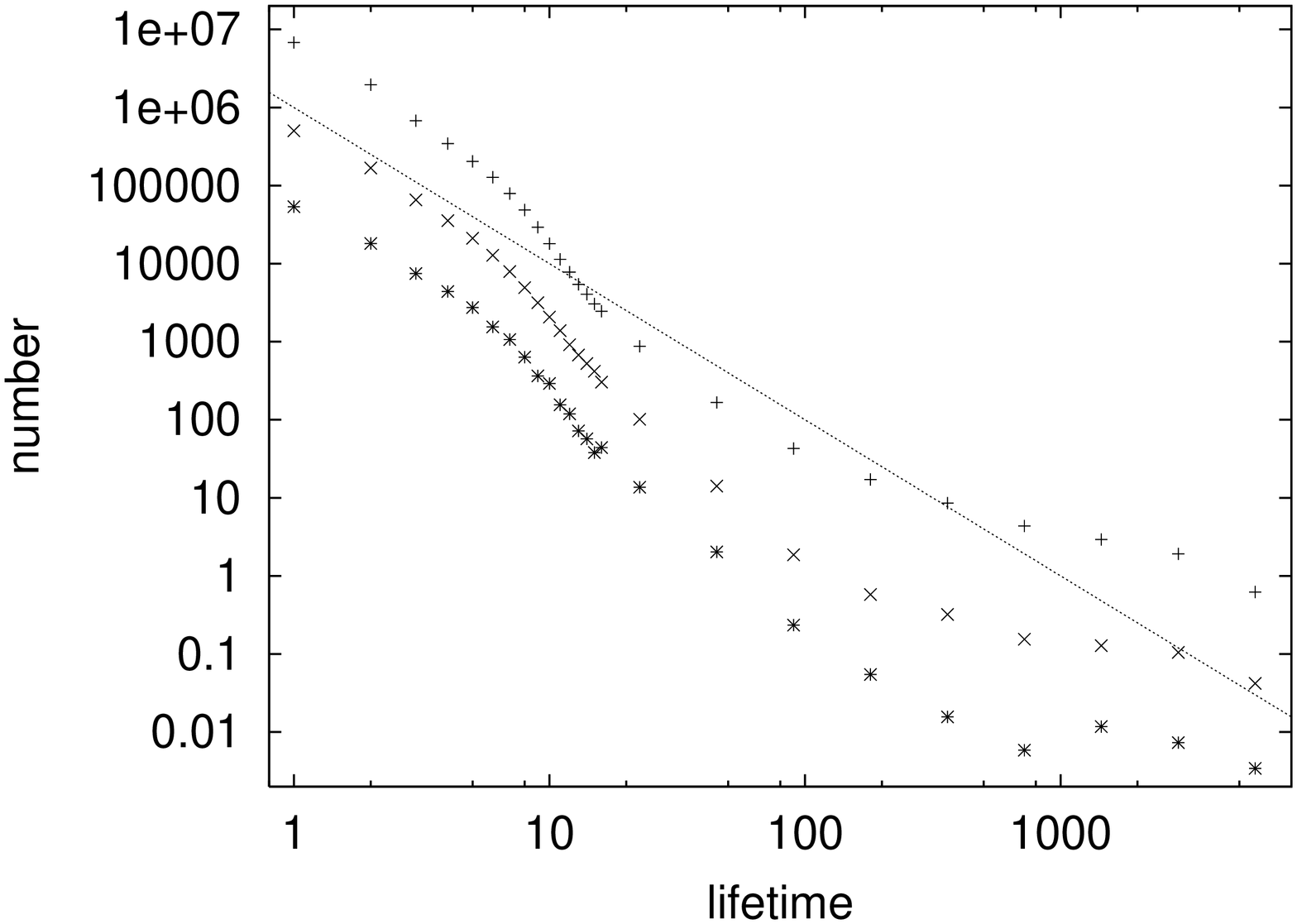}
\includegraphics[angle=-90,width=0.9\columnwidth]{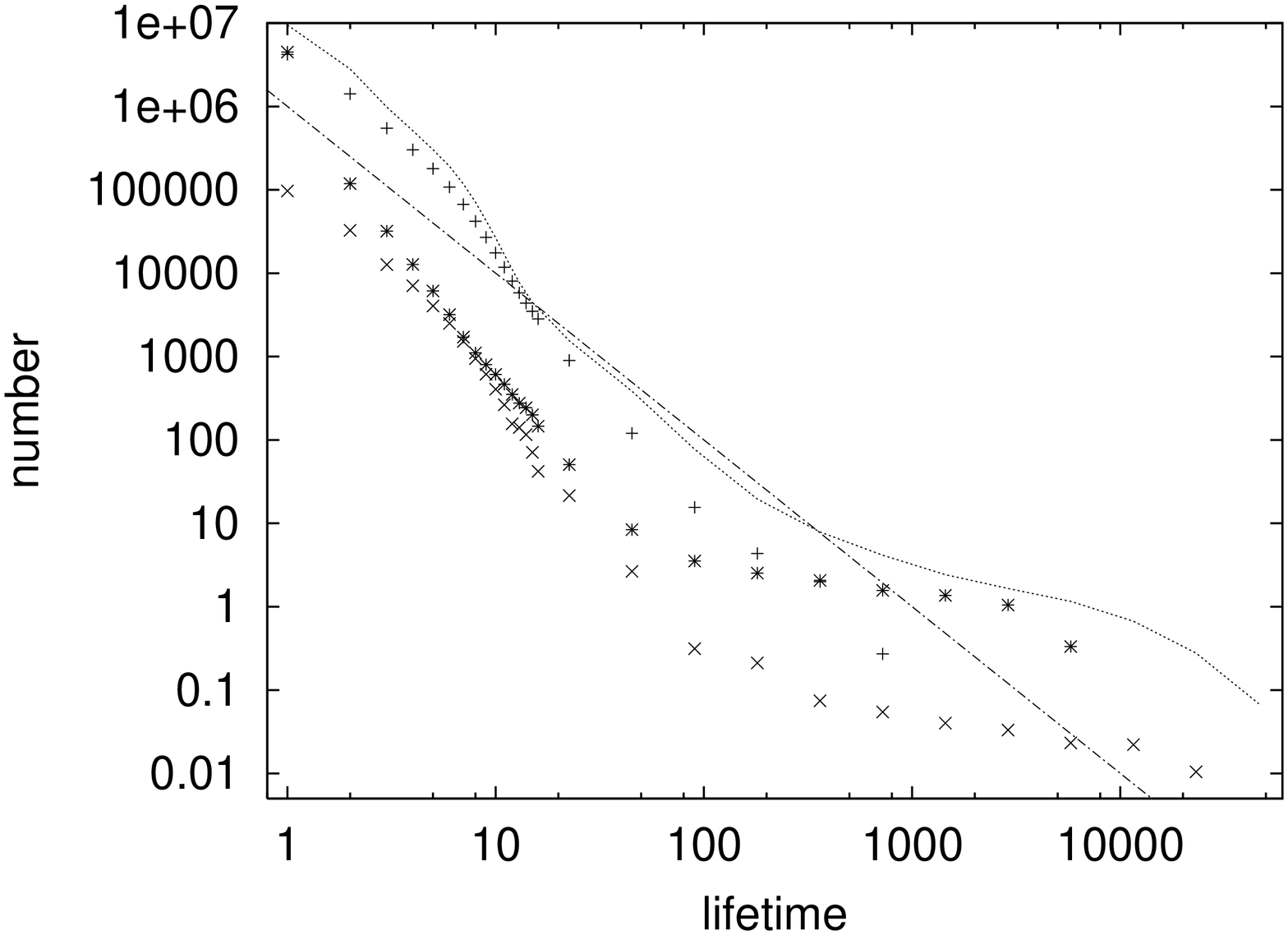}
\end{center}
\caption{Log-log plots of the distributions of the lifetimes of the 
species in an eco-system with $n_{max} = 10^2$ to $10^4$ and 600 to
60000 iterations.  
The line with slope $-2$ corresponds to a power law distribution 
that has been predicted by many theories. 
The common parameters for both plots are 
$m=2, {\ell} = 5$ (i.e. $N_{max} = 31), p_{sp} = 0.1, p_{mut} = 0.0001, 
C = 0.05$). In the upper plot, the symbols $ +, \times$ and $\ast$ correspond to
$n_{max} = 10^2, 10^3, 10^4$ averaged over 6400, 640 and 64 systems
respectively. In the lower plot, $n_{max} = 1000$ (except for the line where 
$n_{max} = 100$) and the maximum simulation 
time is 600 (+) and 60000 ($\times$ and line) iterations; $\ast$
corresponds to $m=12, \ell = 3$ after 6000 iterations; 640 systems were 
averaged over for short and intermediate times, and 64 for the longest time.
Each system started from a new random initial state.}
\label{fig-2}
\end{figure}
%%%%%%%%%%%%%%%%%%%%%%%%%%%%%%%%%%%%%%%%%%%%%%%%%%%%%%%%%%%%%%%%%%%%

\subsection{Probability of birth}

We assume the {\it time-dependent} probability $p_b(i,\alpha)$ (of 
individual $\alpha$ in species $i$) of giving birth per unit time  to
decrease linearly with age, from its maximum value, attainable at the 
minimum reproduction age, down to zero at the maximum lifespan. It is 
multiplied with a Verhulst factor $ 1 - n_i/n_{max}$ and equals this 
factor at $X = X_{rep}$. Thus  in the limit of vanishingly small 
population, i.e., $n_i \rightarrow 0$, we have $p_b(i,\alpha) \rightarrow 1$  
if $X(i,\alpha) = X_{rep}(i)$ and, thereafter, $p_b$ decreases linearly  
\cite{austad} as the organism grows older. However, since the eco-system 
can support only a maximum of $n_{max}$ individual organisms of each 
species, $p_b(i,\alpha;t) \rightarrow 0$ as $n_i(t) \rightarrow n_{max}$, 
irrespective of the age of the individual organism $\alpha$ \cite{cebrat}.

\subsection{Probability of natural death}
Similarly, we assume the probability $p_d$ of ``natural'' death (due to 
ageing) to increase linearly with age \cite{carey} and to reach unity at 
the maximum lifespan $X_{max}$ of the species: 
$p_d = (X M - X_{rep})/(X_{max}M -X_{rep})$.
(For $X < X_{rep}$ the death probability, instead, has the constant 
value that $p_d$ attains at $X = X_{rep}$; if the above denominator is 
negative, $p_d=1$.) Note that, for a given $X_{max}$ and $X_{rep}$, 
the larger is the $M$ the higher is the $p_d$ for any age $X$. 
Therefore, each species has a tendency to increase $M$ for giving  
birth to larger number of offsprings whereas the higher mortality 
for higher $M$ opposes this tendency \cite{tradeoff}.

\section{Results} 

In our simulations. initially, $M = 10, \; X_{max}$ is distributed 
randomly between $2$ and $99$ independently for each species, $X_{rep}$ 
randomly between $1$ and $X_{max}$, the population randomly between $1$ 
and $n_{max}/2$. The ages of the individuals vary randomly between $1$ 
and the $X_{max}$ of their species.

The longest runs in our computer simulations were continued upto a 
a million time steps. If each time step in our 
model is assumed to correspond to a real time of the order of one 
year, then the time scale of a million years, over which we have 
monitored our model eco-system, is comparable to real speciation time
scales. 

%%%%%%%%%%%%%%%%%%%%%%%%%%%%%%%%%%%%%%%%%%%%%%%%%%%%%%%%%%%%%%%%%%%
\begin{figure}[tb]
\begin{center}
\includegraphics[angle=-90,width=0.9\columnwidth]{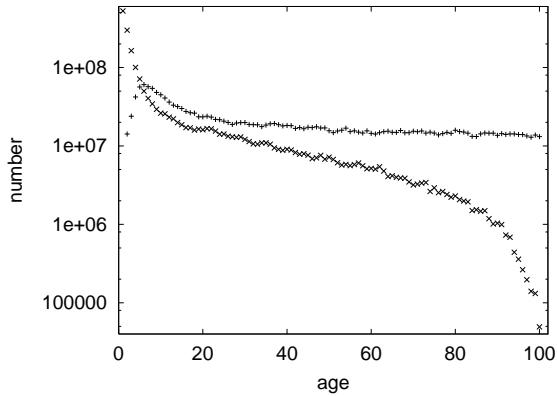}
\end{center}
\caption{Semi-log plot of the distributions of $X_{rep}$ ~($\times$) and $X_{max}$, taken from the simulations symbolized by the curved line in the lower
part of fig.\ref{fig-1}: $m=2, \ell=5, n_{max}=100, t=60000$, 640 systems. 
}
\label{fig-3}
\end{figure}
%%%%%%%%%%%%%%%%%%%%%%%%%%%%%%%%%%%%%%%%%%%%%%%%%%%%%%%%%%%%%%%%%%%%

\subsection{Lifetime distributions}

The average distributions of the lifetimes of the species are plotted 
in fig.\ref{fig-2} for various sets of values of the parameters. Only very
approximately,  
the data are consistent with a power-law; the effective exponent, which 
is, ~~approximately, $2$, is also consistent with the corresponding estimate 
quoted in the literature \cite{drossel,newman}. However, in 
fig.\ref{fig-2} the power law holds only over a limited range \cite{chu} of
times; for longer times a plateau seems to develop. 
Since real eco-systems  are much more complex than our model eco-system 
and the available fossil data are quite sparse, it is questionable 
whether real extinctions follow power laws and, if so, over how many 
orders of magnitude.

%%%%%%%%%%%%%%%%%%%%%%%%%%%%%%%%%%%%%%%%%%%%%%%%%%%%%%%%%%%%%%%%%%%%
\begin{figure}[tb]
\begin{center}
\includegraphics[angle=-90,width=0.9\columnwidth]{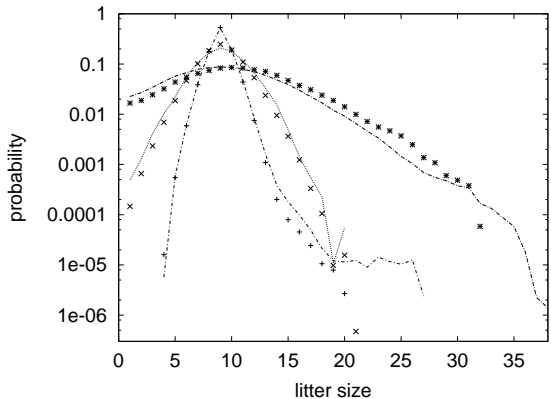}
\end{center}
\caption{Semi-log plot of the distribution of $M$. The parameter 
values are same as those in fig.\ref{fig-3}; shorter and longer simulations  
are added to show further broadening of the distribution. 
The symbols $+$, $\times$ and $\ast$ correspond to $600$, $60000$ and $600000$ 
iterations, respectively, using 6400, 640 and 1 systems. The lower lines, using
64 lattices with $n_{max}=100, \; t=6000$, show the broadening with increasing
mutation rate $p_{mut} = 0.00001, \; 0.001$ and 0.01.
}
\label{fig-4}
\end{figure}
%%%%%%%%%%%%%%%%%%%%%%%%%%%%%%%%%%%%%%%%%%%%%%%%%%%%%%%%%%%%%%%%%%% 

%%%%%%%%%%%%%%%%%%%%%%%%%%%%%%%%%%%%%%%%%%%%%%%%%%%%%%%%%%%%%%%%%%%%%
\begin{figure}[tb]
\begin{center}
\includegraphics[angle=-90,width=0.9\columnwidth]{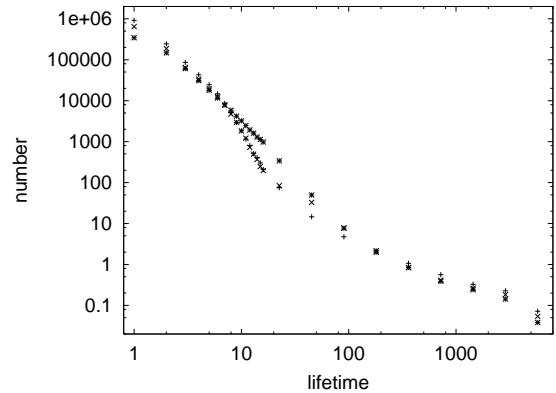}
\end{center}
\caption{Log-log plot of the distribution of lifetimes for speciation
probabilities $p_{sp}$ = 0.02 (+) and  0.5 ($\times$), and (squares, with 
$p_{sp}$ = 0.1) for 
Gompertz mortality assumption: $p_d = \exp[(\max(X,X_{rep})-X_{max})/M]$, 
using 640 systems for $n_{max}=100$ and $t=6000$.}
\label{fig-5}
\end{figure}
%%%%%%%%%%%%%%%%%%%%%%%%%%%%%%%%%%%%%%%%%%%%%%%%%%%%%%%%%%%%%%%%%%%%

%%%%%%%%%%%%%%%%%%%%%%%%%%%%%%%%%%%%%%%%%%%%%%%%%%%%%%%%%%%%%%%%%%%%
\begin{figure}[tb]
\begin{center}
\includegraphics[angle=-90,width=0.9\columnwidth]{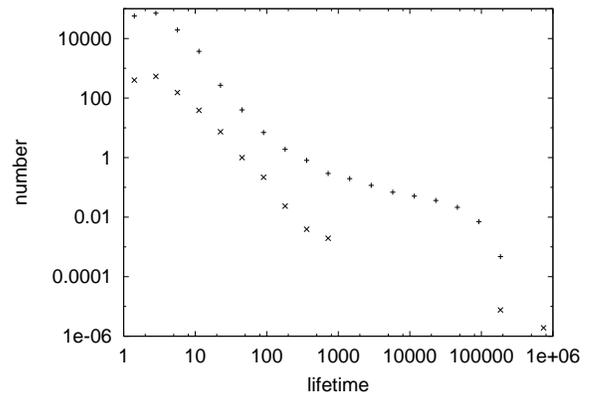}
\end{center}
\caption{Log-log plot of the distribution of lifetimes for the whole 
ecosystem of $\ell$ trophic layers, with $\ell=5,m=2$ and $\ell=3,m=6$,
from 1 and 10 systems only; $n_{max}=100$.
}
\label{fig-6}
\end{figure}
%%%%%%%%%%%%%%%%%%%%%%%%%%%%%%%%%%%%%%%%%%%%%%%%%%%%%%%%%%%%%%%%%%%%

\subsection{Distributions of Species Characteristics}

Figs.\ref{fig-3},\ref{fig-4} show the time-averaged distributions of 
$X_{max}$, $X_{rep}$ and $M$. We see that the minimum age of reproduction 
$X_{rep}$ is quite small, as usual in a similar ageing model 
\cite{stauffer}. The age distribution (not shown) decays stronger than a simple
exponential, indicating a mortality increasing with age as it should 
be \cite{carey}. The genetic death ages $5 < X_{max} < 100$ reach ages far 
above the upper end $\simeq 50$ of the age distribution (for the species on top
of the food web), as is
appropriate for animals in the wild \cite{austad}. Finally, 
fig.\ref{fig-4} shows the distribution of $M(i)$ which is still broadening
even after 60000 iterations.

We have also observed (not shown)
that the higher is the mutation probability $p_{mut}$ the lower 
is the lifetime of the eco-system; 
this is consistent with the intuitive expectation that 
a higher rate of mutation leads to higher levels of biological activity 
in the eco-system thereby leading to the extinction of larger number of species.
% new remark added below; also figure capotion 4 corrected and expanded
Fig.\ref{fig-4} from these data shows that the broadening of the histogram
for $M$, i.e. the equilibration process, is determined by the product $p_{mut}t$
giving the average number of mutations per species. 
But, $p_{sp}$ had weaker effect on the same data as shown in 
Fig. \ref{fig-5}. The same figure also shows a somewhat better power law 
at short times if the above linear increase of the mortality with age
is replaced by an exponential increase (Gompertz law \cite{austad}).   

\subsection{Collapse of fragile ecosystems}

We model an eco-system with a {\it fixed} number $\ell$ of trophic levels; 
thus as soon as we find one level to be extinct completely, we regard the
eco-system as destroyed and try to build a new one for the same parameters,
changing only the random numbers. Hundreds of such attempts are needed for
a successful system lasting the prescribed number (like 6000) of iterations,
see fig.\ref{fig-6}. 
This method simulates the billions of years which natural evolution 
needed to build the present life on earth. 

\section{Summary and conclusion}

In summary, we have presented a unified model which describes not only 
the birth, ageing and death of individuals as well as population dynamics 
on short time scales but also the long-time evolution of species, their 
origination/speciation and extinction. The total number of species, the 
inter-species interactions and the collective characteristics, namely, 
$X_{rep}, X_{max}$ and $M$, of each species vary following a stochastic 
dynamics with Darwinian selection. Thus, our model is capable of 
{\it self-organization}. 

\noindent {\bf Acknowledgements} 

We thank J.E. Cohen for emphasizing to us the importance of food webs and the 
Supercomputer Center J\"ulich for computer time on their CRAY-T3E.
This work is supported by Deutsche Forschungsgemeinschaft through a 
Indo-German joint research project. 

\bibliographystyle{plain}

\end{document}